\documentclass[%
reprint,
superscriptaddress,
 amsmath,amssymb,
 aps,
pra,twocolumn,
]{revtex4}

\usepackage{graphicx}
\usepackage{dcolumn}
\usepackage{bm}
\usepackage{hyperref}
\usepackage{amsthm}
\usepackage{amsfonts}
\usepackage{mathrsfs}
\theoremstyle{definition}

\newtheorem{theorem}{Theorem}

\newcommand{\bra}[1]{\langle#1|}
\newcommand{\ket}[1]{|#1\rangle}
\newcommand{\braket}[2]{\langle#1|#2\rangle}
\newcommand{\ketbra}[2]{|#1\rangle\!\langle#2|}

\begin{document}


\title{Limits on inference of gravitational entanglement}

\author{Yue Ma}
\affiliation{QOLS, Blackett Laboratory, Imperial College London, London SW7 2AZ, United Kingdom}

\author{Thomas Guff}
\affiliation{Department of Physics, Stockholm University, AlbaNova University Center, SE-106 91 Stockholm, Sweden}

\author{Gavin W. Morley}
\affiliation{Department of Physics, University of Warwick, Coventry, CV4 7AL, United Kingdom}

\author{Igor Pikovski}
\affiliation{Department of Physics, Stockholm University, AlbaNova University Center, SE-106 91 Stockholm, Sweden}
\affiliation{Department of Physics, Stevens Institute of Technology, Castle Point on the Hudson, Hoboken, NJ 07030, USA}

\author{M. S. Kim}
\affiliation{QOLS, Blackett Laboratory, Imperial College London, London SW7 2AZ, United Kingdom}



\begin{abstract}
Combining gravity with quantum mechanics remains one of the biggest challenges of physics. In the past years, experiments with opto-mechanical systems have been proposed that may give indirect clues about the quantum nature of gravity.
In a recent variation of such tests [D. Carney et al., Phys.Rev.X Quantum 2, 030330 (2021)], the authors propose to gravitationally entangle an atom interferometer with a mesoscopic oscillator. The interaction results in periodic drops and revivals of the interferometeric visibility, which under specific assumptions indicate the gravitational generation of entanglement. Here we study semi-classical models of the atom interferometer that can reproduce the same effect. We show that the core signature -- periodic collapses and revivals of the visibility -- can appear if the atom is subject to a random unitary channel, including the case where the oscillator is fully classical and situations even without explicit modelling of the oscillator. We also show that the non-classicality of the oscillator vanishes unless the system is very close to its ground state, and even when the system is in the ground state, the non-classicality is limited by the coupling strength. Our results thus indicate that deducing entanglement from the proposed experiment is very challenging, since fulfilling and verifying the non-classicality assumptions is a significant challenge on its own right.  

\end{abstract}

\maketitle

\section{Introduction}
The search for a full theory of quantum gravity is a major open problem in modern physics. The difficulty to find such a theory has even raised conceptual questions on the need for the quantization of gravity \cite{eppley1977necessity, penrose1996gravity, dyson2014graviton, baym2009two, mari2016experiments, belenchia2018quantum, rydving2021gedankenexperiments, carney2021newton}. One major challenge is the lack of experimental evidence. However, in recent years experimental proposals to test the quantization of gravity have become an active research field. On the one hand, quantum gravity phenomenology offers alternative models that can be probed from astrophysical observations \cite{amelino1998tests, hossenfelder2017experimental} and table-top experiments \cite{marshall2003towards, pikovski2012probing, bekenstein2012tabletop, bassi2017gravitational}. On the other hand, the entanglement between two gravitating systems can provide indirect signatures of quantized gravity \cite{bose2017spin, marletto2017gravitationally}. In similar spirit as the latter, a recent paper by Carney, M\"{u}ller and Taylor ~\cite{Carney2021} showed that interactions between atoms and a massive systems can hint at the quantum nature of gravity. They showed that the coupling results in periodic collapses and revivals of the interferometer visibility of the atomic interferometer. The authors show that under specific conditions such as Markovianity and time independent Hamiltonians, this behavior implies entanglement between the atom and the harmonic oscillator through gravity, hence to conclude the quantum nature of gravity.

Here we study classical models for the loss and revival of visibility. Our analysis shows that this signature can be reproduced if the atom evolves according to a random unitary channel, without being coupled to another quantum system~\cite{audenaert2008random}. As the central part of the atomic interferometer is the accumulated phase during the time evolution, the idea is motivated by a previous work~\cite{armata2016quantum} where the optical phase originating from an optomechanical interaction is found to have a classical origin, such that classical models can explain supposed quantum behaviour in other proposals, such as in Ref. \cite{marshall2003towards}. For the case where both the atoms and the oscillator are described fully quantum mechanically, we study the non-classicality for a thermal harmonic oscillator, showing that it vanishes for low coupling even if the system is in the ground state. Therefore, such experiments with very low coupling strengths and at finite temperature always allow for a classical description, unless it is explicitly invalidated experimentally.


\section{the original quantum model}\label{sec:level1}

The setup described in Ref.~\cite{Carney2021} consists of an atom localized into one of two positions interacting with a quantum harmonic oscillator. The position degree of freedom of the atom can therefore be represented as a qubit. 
The atom will interact gravitationally with the harmonic oscilltor according to the Hamiltonian
\begin{equation}\label{eq:qHam}
    \hat{H}=\omega\hat{a}^{\dagger}\hat{a}+g(\hat{a}+\hat{a}^{\dagger})\hat{\sigma}_z.
\end{equation}
Here $\hat{a}$ and $\hat{a}^{\dagger}$ are annihilation and creation operators for the mechanical oscillator. They are related to the position and momentum operators of the mechanical oscillator via $\hat{a}=\sqrt{m\omega/2}\hat{X}+i/\sqrt{2m\omega}\hat{P}$, where $m$ and $\omega$ are the mass and frequency of the oscillator, respectively. The operator $\hat{\sigma}_z$ acts on the atom, defined as
\begin{equation}
    \hat{\sigma}_z=|1\rangle\langle1|-|0\rangle\langle0|,
\end{equation}
where $|0\rangle$ and $|1\rangle$ are two states of the position degree of freedom of the atom. 
The coupling strength $g$ depends on the gravitational force between the two systems (i.e. the masses of the two systems and the distance between them).
Note that throughout this article we have set $\hbar=1$.

The mechanical oscillator is initially in a thermal state at temperature $T$, described by
\begin{equation}
    \rho_{\mathrm{th}}=\frac{1}{\pi\tilde{n}}\int d^2\alpha\; e^{-\frac{|\alpha|^2}{\tilde{n}}} \ketbra{\alpha}{\alpha}, \label{eq:thermal}
\end{equation}
where $\tilde{n}=(\exp(\omega/k_{\mathrm{B}}T)-1)^{-1}$ is the thermal phonon number and $\ket{\alpha}$ is a coherent state. The atom is initially in the state $|0\rangle$.

The experimental proposal in \cite{Carney2021} is to then perform  interferometry on the atom.
This consists of the following steps. The Hadamard gate is applied to the atom, resulting in the transformation
\begin{subequations}\label{eq:hadamard}
\begin{align}
    &|0\rangle\rightarrow \ket{+}=\frac{1}{\sqrt{2}}(|0\rangle+|1\rangle),\\
    &|1\rangle\rightarrow\ket{-}=\frac{1}{\sqrt{2}}(|0\rangle-|1\rangle).
\end{align}
\end{subequations}
Then the atom-oscillator system evolves according to the Hamiltonian Eq.~\eqref{eq:qHam} for time $t$, described by the unitary operator
\begin{equation}
    \hat{U}_q(t)=e^{-i\hat{H}t}.
\end{equation}
To describe the system in a thermal state, we first calculate the evolution for an arbitrary coherent state $\ket{\alpha}$.
After evolving for a time $t$ the combined atom-oscillator state is given by (up to a global phase),
\begin{align}
    \hat{U}_q(t)&\ket{+}\ket{\alpha} \nonumber \\
    &= \frac{1}{\sqrt{2}}\left(e^{i\theta(t)} \ket{0}\ket{\alpha_{+}(t)} + e^{-i\theta(t)}\ket{1}\ket{\alpha_{-}}\right)
\end{align}
where
\begin{subequations}
\begin{align}
\alpha_{\pm}(t) &= \alpha e^{-i\omega t}\pm \frac{g}{\omega}\left(1-e^{-i\omega t}\right), \\
\theta(t) &= \frac{g}{\omega}\text{Im}(\alpha(1-e^{-i\omega t})).
\end{align}
\end{subequations}
Therefore, if the oscillator begins in a thermal state, the combined system will evolve under $\hat{U}_{q}(t)$ to the state
\begin{align}
    &\hat{U}_q(t)\left(\ketbra{+}{+}\otimes \rho_{\text{th}}\right)U^{\dagger}_{q}(t) \nonumber \\
    &\;= \frac{1}{\pi\tilde{n}} \int e^{-\frac{|\alpha|^2}{\tilde{n}}}\frac{1}{2}\left(e^{i\theta(t)} \ket{0}\ket{\alpha_{+}(t)} + e^{-i\theta(t)}\ket{1}\ket{\alpha_{-}}\right) \nonumber \\
    &\qquad\qquad \times \left(e^{-i\theta(t)} \bra{0}\bra{\alpha_{+}(t)} + e^{i\theta(t)}\bra{1}\bra{\alpha_{-}}\right)d^{2}\alpha. \label{eq:evolve}
\end{align}
After this evolution a phase shift $\varphi$ is applied to the atom state $|1\rangle$, which is realised by the unitary operator
\begin{equation}
    \hat{U}_{\varphi}=e^{i\varphi}|1\rangle\langle1|+|0\rangle\langle0|.
\end{equation}
Then another Hadamard gate Eq.~\eqref{eq:hadamard} is applied to the atom. Finally the position of the atom is measured. The probability of the atom to be in the state $|0\rangle$ is
\begin{equation}\label{eq:probQ}
    P_{q,\varphi}=\frac{1}{2}+\frac{1}{2}e^{-16\frac{g^2}{\omega^2}(\tilde{n}+\frac{1}{2})\sin^2\frac{\omega t}{2}}\cos\varphi.
\end{equation}
The interference visibility is defined as
\begin{equation}
    V=\frac{\max_{\varphi}P_{q,\varphi}-\min_{\varphi}P_{q,\varphi}}{\max_{\varphi}P_{q,\varphi}+\min_{\varphi}P_{q,\varphi}}.
\end{equation}
After inserting in Eq.~\eqref{eq:probQ} we get
\begin{equation}\label{eq:qV}
    V=e^{-16\frac{g^2}{\omega^2}(\tilde{n}+\frac{1}{2})\sin^2\frac{\omega t}{2}}.
\end{equation}
The visibility decays to its minimum value at $t=\pi/\omega$,  when the atom is maximally  entangled with the harmonic oscillator \eqref{eq:evolve}. The visibility then returns to the maximum value $1$ at $t=2\pi/\omega$, when the atom is fully disentangled from the harmonic oscillator. This pattern is repeated with the period $2\pi/\omega$. These are referred to as the periodic collapse and revival of the interference visibility, which in the fully quantum mechanical picture can be attributed to the entanglement between the two systems.

The periodic appearance and disappearance of entanglement can be clearly seen by examining the explicit form of the unitary time evolution operator. For this purpose, we generalise the $\hat{\sigma}_z$ operator in Eq.~\eqref{eq:qHam} to any operator $\hat{O}$ that commutes with $\hat{a}$ and $\hat{a}^{\dagger}$,
\begin{equation}\label{eq:qGen}
    \hat{H}_O = \omega\hat{a}^{\dagger}\hat{a}+g(\hat{a}+\hat{a}^{\dagger})\hat{O}.
\end{equation}
The corresponding unitary time evolution operator can be expressed as
\begin{align}\label{eq:UO}
    &\hat{U}_O(t)=e^{-i\hat{H}_Ot}\\
    &=e^{i\frac{g^2}{\omega^2}(\omega t-\sin\omega t)\hat{O}^2}e^{-i\omega t\hat{a}^{\dagger}\hat{a}}e^{-\frac{g}{\omega}((e^{i\omega t}-1)\hat{a}^{\dagger}-(e^{-i\omega t}-1)\hat{a})\hat{O}}.\nonumber
\end{align}
The last exponential factor is a displacement operator of the oscillator, conditioned on the state of the atom.
At times $t=2n\pi/\omega$ where $n\in\mathbb{Z}$, the last two factors in Eq.~\eqref{eq:UO} reduce to the identity, meaning that at these times $\hat{U}_O(t)$ is independent of $\hat{a}$ and $\hat{a}^{\dagger}$, and the atom and the oscillator decouple. The first factor is usually associated with a nonlinear geometric phase gate on the mode described by $\hat{O}$. In \cite{armata2016quantum}, a similar interferometric setup was considered but with coherent states of light. In that case the non-linear factor in \eqref{eq:UO} caused an additional loss of visibility. Since the model considered here only studies interferometry with a single qubit, there is no additional loss of visibility. Indeed since $\hat{\sigma}_z^2$ is the identity operator, the non-linear factor only appears as a global phase.

\section{Semi-classical Approaches}

In this section we present several semi-classical models which reproduce the same periodic collapse and revival pattern as seen in the fully quantum mechanical case \eqref{eq:qV}. In these models, the only quantum element in the setup is the atom, which is modelled as a two-level system. We start with a general formalism where the atom is subject to a random unitary channel, then we explicitly consider three examples.

\subsection{General Formalism}

In our semi-classical models we perform the same atom interferometry experiment but we replace $\hat{U}_q(t)$ with a random unitary channel. To be specific, the atom is prepared in the $|+\rangle$ state. It then evolves under a random unitary channel whose effect on the atomic state is described by
\begin{equation}\label{eq:scState}
    \rho(t)=\langle \hat{U}_{sc}(t)|+\rangle\langle+|\hat{U}_{sc}^{\dagger}(t) \rangle_c,
\end{equation}
where $\hat{U}_{sc}(t)$ a phase shift unitary operator,
\begin{equation}\label{eq:scU}
    \hat{U}_{sc}(t)=e^{i\phi(t)}|1\rangle\langle1|+|0\rangle\langle0|,
\end{equation}
$\phi(t)$ is a real-valued random variable at each time, and $\langle \cdot \rangle_c$ refers to taking the average over the classical probability distribution of random variables. The phase shift $\phi(t)$ can, for instance, be generated via the Hamiltonian of the atom,
\begin{equation}
    \hat{H}_{sc}=G(t)\hat{\sigma}_z,\label{eq:HamGen}
\end{equation}
where
\begin{equation}
    G(t)=-\frac{1}{2}\frac{d\phi(t)}{dt}.\label{eq:gT}
\end{equation}
The atomic state can be explicitly written as
\begin{align}
    \rho(t)&=\frac{1}{2}|1\rangle\langle 1|+\frac{1}{2}|0\rangle\langle 0|+\frac{1}{2}\langle e^{i\phi(t)} \rangle_c|1\rangle\langle0|\nonumber\\
    &+\frac{1}{2}\langle e^{-i\phi(t)} \rangle_c|0\rangle\langle1|.
\end{align}
To finish the interferometry, same as the quantum case, a phase shift $\varphi$ is applied to $|1\rangle$, then the Hadamard gate Eq.~\eqref{eq:hadamard} acts on the atom, and finally we measure the atom position. The interference visibility is derived to be
\begin{equation}
    V=|\langle e^{i\phi(t)} \rangle_c|.
\end{equation}
The condition for reproducing the quantum collapse and revival of the visibility, is therefore
\begin{equation}\label{eq:condition}
    |\langle e^{i\phi(t)} \rangle_c| = e^{-16\frac{g^2}{\omega^2}(\tilde{n}+\frac{1}{2})\sin^2\frac{\omega t}{2}}.
\end{equation}
Any random unitary channel, i.e., classical probability distribution of $\phi(t)$, which can satisfy the condition Eq.~\eqref{eq:condition}, will reproduce the same visibility as a function of time as governed by the quantum interaction Hamiltonian Eq.~\eqref{eq:qHam}. 
This will be true if the classical uncertainty associated with $\phi(t)$ vanishes with period $2\pi/\omega$, but remains at intermediate times.
We will now give several examples of such a $\phi(t)$.

\subsection{Example semi-classical model 1} \label{sec:ex1}

The first semi-classical model is based on a semi-classical mean-field approximation to the quantum Hamiltonian Eq.~\eqref{eq:qHam}. We assume that the atom is a two-state quantum system, while the mechanical oscillator is classical. The atom evolves according to the Hamiltonian
\begin{equation}\label{eq:sc1h}
    \hat{H}_{sc1}=\sqrt{2}g x(t)\hat{\sigma}_z,
\end{equation}
where $x(t)$ is the dimensionless position of the mechanical oscillator, which is related to the physical displacement $X$ of the oscillator via the `zero-point length', $x=X\sqrt{m\omega}$. The mechanical oscillator is assumed to only see the mean-field effect of the atom, i.e., the force applied from the atom onto the mechanical oscillator is $F=-\sqrt{2m\omega}g\langle \hat{\sigma}_z \rangle$. As the atom is in the state $(|0\rangle+|1\rangle)/\sqrt{2}$ before the interaction with the mechanical oscillator starts, and the interaction Eq.~\eqref{eq:sc1h} only induces a phase difference between the two basis states $|0\rangle$ and $|1\rangle$, $F=0$ holds throughout the time evolution. Therefore we can write the time evolution of the dimensionless mechanical oscillator position as
\begin{equation}
    x(t)=x_0\cos\omega t+p_0\sin\omega t,
\end{equation}
where $x_0$ is the initial value of dimensionless position and $p_0\sqrt{m\omega}$ is the initial value of the momentum. The energy of the classical oscillator is therefore
\begin{equation}
    E(x_{0},p_{0}) = \frac{\omega}{2}(x_{0}^{2}+p_{0}^{2}).
\end{equation}
The Hamiltonian \eqref{eq:sc1h} induces evolution according to the unitary
\begin{equation}
    U_{sc1}(t) = \exp\left(-i \sqrt{2} g \int_{0}^{t} d\tau x(\tau) \sigma_{z}\right).
\end{equation}
This is (up to a global phase) of the form \eqref{eq:scU}, where
\begin{align}\label{eq:sc1Phase}
    \phi(t)&=-2\sqrt{2}g\int_0^t d\tau x(\tau)\nonumber\\
    &=-2\sqrt{2}\frac{g}{\omega}(x_{0}\sin\omega t+p_{0}(1-\cos\omega t)).
\end{align}
If we assume the classical harmonic oscillator is in a thermal state at inverse temperature $\beta$, then the energy distribution is given by a Boltzmann distribution,
\begin{equation}
    p(E(x_{0},p_{0})) = \frac{\beta\omega}{2\pi}e^{-\frac{\beta\omega}{2}(x_{0}^{2}+p_{0}^{2})}.
\end{equation}
Thus we see the variables $x_{0}$ and $p_{0}$ have a normal distribution with standard deviation $\sqrt{1/\beta\omega}$.
Therefore, the interferometric visibility is
\begin{align}\label{eq:sc1V}
    \left|\langle e^{i\phi(t)} \rangle_c \right| &= \left|\int dx_{0}dp_{0}\, p(E(x_{0},p_{0}))\, e^{i\phi(t)}\right| \nonumber \\
    &=e^{-16n_c\frac{g^2}{\omega^2}\sin^2\frac{\omega t}{2}},
\end{align}
where we defined the classical phonon number as $n_c=1/\beta\omega$.
This is in the same form as Eq.~\eqref{eq:condition} except that, the factor $\tilde{n}+1/2$ in the exponential is replaced by $n_c$. Recall that the quantum thermal phonon number is expressed as $\tilde{n}=1/(\exp(\omega/k_{\mathrm{B}}T)-1)$. For high temperature, $k_{\mathrm{B}}T\gg \omega$,  we have that $\tilde{n}+1/2 \approx n_c$. The visibility Eq.~\eqref{eq:sc1V} is thus indistinguishable from the quantum visibility Eq.~\eqref{eq:qV}. If the temperature is low, the difference between $\tilde{n}+1/2$ and $n_c$ is significant. However, the quantum visibility can be reproduced if the classical oscillator begins in a higher temperature, so that the standard deviations of $x_0$ and $p_0$ are proportional to $\sqrt{\tilde{n}+1/2}$, instead of the thermal width $\sqrt{n_c}$. 

We can explain the revival and collapse of the interferometric visibility in this semi-classical model as follows. The phase shift in this example is proportional to the integral of the position \eqref{eq:sc1Phase} of the classical oscillator from its initial position at time $t=0$. The uncertainty in the initial position and momentum of the oscillator translates into uncertainty into the phase shift which leads to a reduction in visibility. However since the motion of the mechanical oscillator is periodic, the integral of the position will be zero with certainty every mechanical period $t=2\pi/\omega$, implying that the phase shift at these times is certainly zero and therefore the visibility will periodically revive.

Note that this semi-classical model based on the mean-field interaction with a classical oscillator shares the same idea as the optomechanical example in Ref.~\cite{armata2016quantum}, and it has been suggested recently~\cite{hosten2021testing} to claim against the proposal in Ref.~\cite{Carney2021}. In the next subsections we describe other semi-classical models that do not correspond to the interaction of the atom with a classical thermal mechanical oscillator, but which nevertheless reproduce the same interference visibility as Eq.~\eqref{eq:qV}.

\subsection{Example semi-classical model 2}

The second semi-classical model assumes an interaction Hamiltonian
\begin{equation}\label{eq:Hsc2}
    \hat{H}_{sc2}=\sqrt{2}g\tilde{x}_0\cos\left(\frac{\omega t}{2}\right)\hat{\sigma}_z,
\end{equation}
where $\tilde{x}_0$ is a Gaussian random variable with mean $0$ and standard deviation $\sqrt{\tilde{n}+1/2}$. The corresponding phase modulation $\phi(t)$ in Eq.~\eqref{eq:scU} is
\begin{equation}\label{eq:sc2Phase}
    \phi(t)=-4\sqrt{2}\frac{g}{\omega}\sin\left(\frac{\omega t}{2}\right)\tilde{x}_0.
\end{equation}
At times $t=2n\pi/\omega$ where $n\in \mathbb{Z}$, $\phi(t)=0$, thus the randomness which depends on $\tilde{x}_0$ disappears, leading to full revival of the interference visibility. It is straightforward to show that the visibility Eq.~\eqref{eq:qV} is recovered.

This example might be considered a special case of the first, with $p_{0}=0$. In such a case there is no uncertainty in the initial momentum of the classical oscillator. As a result, the integral of the position \eqref{eq:sc1Phase} is periodically zero with half the period, therefore in order to match \eqref{eq:qV} in this example, the classical oscillator must have half the frequency, as seen in \eqref{eq:Hsc2}.

The phase modulation Eq.~\eqref{eq:sc2Phase} is the product of one random variable and a periodic time-dependent function. In comparison, in the first semi-classical model, Eq.~\eqref{eq:sc1Phase} contains two random variables, each one multiplied by a periodic time-dependent function. It is possible to construct more semi-classical models, by summing up larger numbers of terms, each term made of the product between a random variable and a periodic time-dependent function. In the next subsection, we will describe a systematic way of constructing semi-classical models based on the characteristic function of classical random variables.

\subsection{Characteristic function method and example semi-classical model 3}

We can construct semi-classical models directly from Eq.~\eqref{eq:scU}. The condition for reproducing the quantum visibility, Eq.~\eqref{eq:condition}, is related to the characteristic function of a random variable. At each time $t$, $\phi(t)$ is a random variable. Its characteristic function is
\begin{equation}
    \Psi_{\phi(t)}(k)=\langle e^{ik\phi(t)} \rangle_c.
\end{equation}
The condition Eq.~\eqref{eq:condition} is therefore the requirement
\begin{equation}\label{eq:char}
    |\Psi_{\phi(t)}(k=1)|=e^{-16\frac{g^2}{\omega^2}(\tilde{n}+\frac{1}{2})\sin^2\frac{\omega t}{2}}.
\end{equation}
There are an infinite number of $\Psi_{\phi(t)}(k)$ (and therefore $\phi(t)$), that satisfy Eq.~\eqref{eq:char}. As an example, we choose
\begin{equation}
    \Psi_{\phi(t)}(k)=e^{-16k^2\frac{g^2}{\omega^2}(\tilde{n}+\frac{1}{2})\sin^2\frac{\omega t}{2}}.
    \label{eq:charFun}
\end{equation}
This can be the characteristic function of a single Gaussian random variable, or the sum of several independent Gaussian random variables. For the former case, $\phi(t)$ is a Gaussian random variable with mean $0$ and time-dependent variance
\begin{equation}
    \sigma^2=32\frac{g^2}{\omega^2}(\tilde{n}+\frac{1}{2})\sin^2\left(\frac{\omega t}{2}\right).
\end{equation}
Note that the semi-classical example 2 considered in the previous subsection is included in this situation. For the latter case, we apply this characteristic function method to explicitly construct another semi-classical model, named example 3. To be specific, we can split the characteristic function Eq.~\eqref{eq:charFun} into the product of two exponentials, each one corresponding to the characteristic function of a Gaussian random variable,
\begin{align}
    \Psi_{\phi(t)}(k)&=e^{-16k^2\frac{g^2}{\omega^2}(\tilde{n}+\frac{1}{2})\sin^2\frac{\omega t}{2}\sin^2\omega t}\nonumber\\
    &\times e^{-16k^2\frac{g^2}{\omega^2}(\tilde{n}+\frac{1}{2})\sin^2\frac{\omega t}{2}\cos^2\omega t}.
\end{align}
Thus $\phi(t)=v_1+v_2$ is the sum of two independent zero-mean Gaussian random variables $v_1$ and $v_2$, with variance
\begin{subequations}
\begin{align}
&\sigma^2_{v_1}=32\frac{g^2}{\omega^2}(\tilde{n}+\frac{1}{2})\sin^2\left(\frac{\omega t}{2}\right)\sin^2(\omega t),\\
&\sigma^2_{v_2}=32\frac{g^2}{\omega^2}(\tilde{n}+\frac{1}{2})\sin^2\left(\frac{\omega t}{2}\right)\cos^2(\omega t).
\end{align}
\end{subequations}
These can be realised by choosing
\begin{subequations}
\begin{align}
    &v_1=4\sqrt{2}\frac{g}{\omega}\sin\left(\frac{\omega t}{2}\right)\sin(\omega t)x_1,\\
    &v_2=4\sqrt{2}\frac{g}{\omega}\sin\left(\frac{\omega t}{2}\right)\cos(\omega t)x_2,
\end{align}
\end{subequations}
where $x_1$ and $x_2$ are two independent Gaussian random variables with mean $0$ and standard deviation $\sqrt{\tilde{n}+1/2}$. By making use of Eqs.~\eqref{eq:HamGen} and \eqref{eq:gT}, we get the Hamiltonian
\begin{align}
    \hat{H}_{sc3}=&-\frac{1}{\sqrt{2}}g\Big[\left(3\sin\frac{3\omega t}{2}-\sin\frac{\omega t}{2}\right)x_1\nonumber\\
    &+\left(3\cos\frac{3\omega t}{2}-\cos\frac{\omega t}{2}\right)x_2\Big]\hat{\sigma}_z.
\end{align}

\section{Non-classicality}
So far our approach has been to treat the problem in a semi-classical picture where only the atom position is a quantum mechanical degree of freedom and it is subject to a random unitary channel. In this section we look at non-classicality of the system when it is treated quantum mechanically as a whole (as described in section~\ref{sec:level1}). To analyze if entanglement can be inferred, we calculate the Wigner function negativity of the oscillator state when it interacts with the atomic interferometer. The negativity of the Wigner function is a measure of the non-classicality of a quantum state \cite{Kenfack2004}, and quantifies the extent to which the corresponding Wigner function has negative values.
The Wigner function of a quantum state $\ket{\psi}$ is defineded as
\begin{equation}
    W(q,p) = \frac{1}{2\pi \hbar} \int dx \braket{q-\frac{1}{2}x}{\psi}\braket{\psi}{q+\frac{1}{2}x}e^{\frac{ipx}{\hbar}}.
\end{equation}
The negativity of a Wigner function is defined as
\begin{align}
    \delta(W) &= \int dq \, dp\, \left(|W(q,p)| - W(q,p)\right) \nonumber \\
    &= \int dq \, dp\, |W(q,p)| - 1. \label{eq:wigneg}
\end{align}

The scheme in \cite{Carney2021} relies on enhancing the sensitivity for detecting entanglement (by witnessing the decline and revival of the entanglement visibility) through increased temperature of the oscillator. To the extent the state \eqref{eq:evolve} is entangled, the state of the oscillator should be in a superposition and hence non-classical. 
The Wigner negativity was used in \cite{Kleckner2008} to study whether an oscillator achieved non-classical states if it is located in one arm of a single photon interferometer. There it was found that the negativity decreased as the initial temperature increased. We do a similar calculation here. 


\begin{figure}[t]
    \centering
    \includegraphics[scale=0.6]{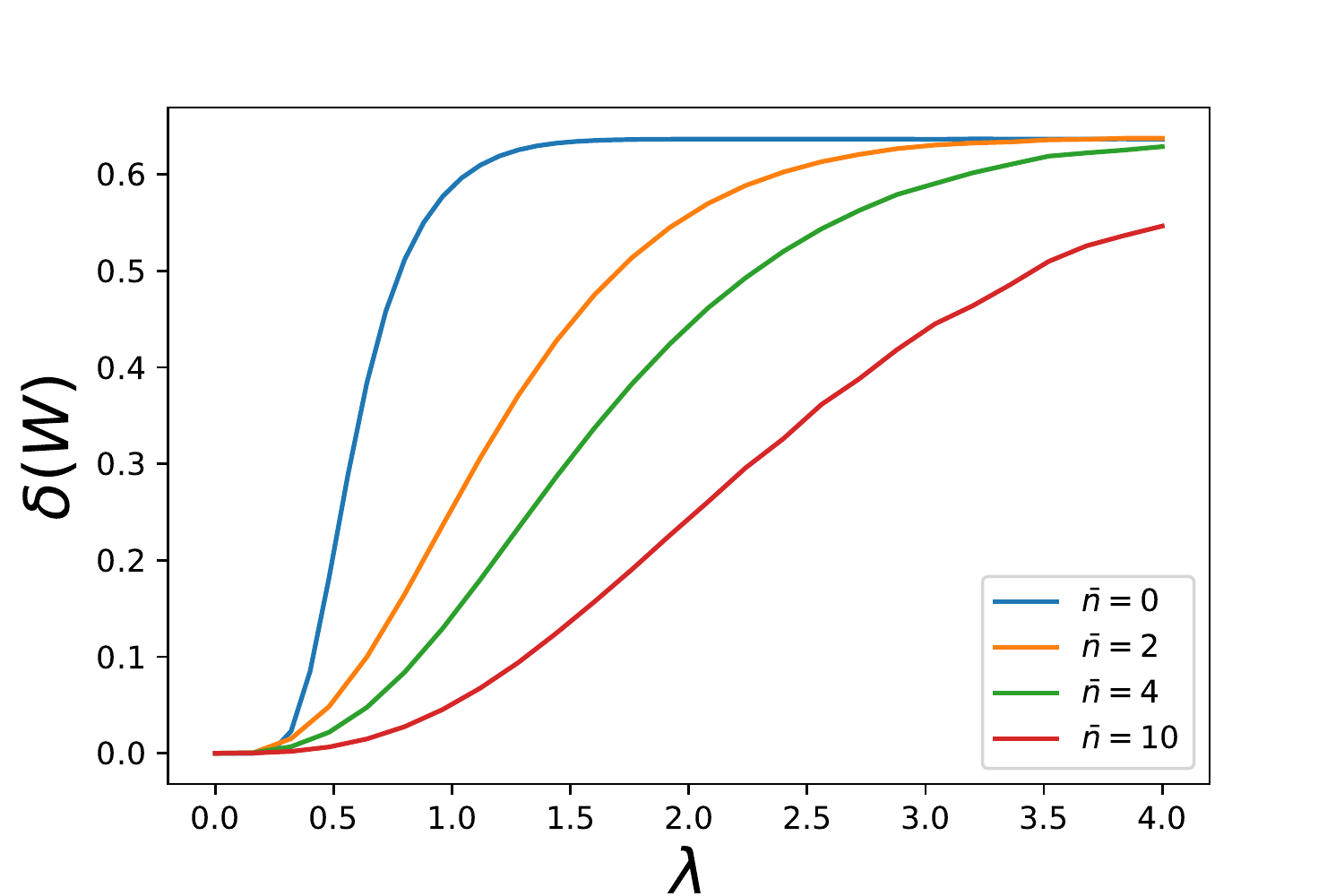}
    \caption{Wigner negativity for the oscillator in state \eqref{eq:evolve} at a half mechanical period $t=\pi/\omega$ and after decoupling from the atom. Wigner function is given by \eqref{eq:wig}. The negativity increases for larger interaction strength $\lambda$ but decreases with increasing initial oscillator temperature.}
    \label{fig:wigneg}
\end{figure}

To study the state of the oscillator directly, we consider it at a half mechanical period $t=\pi/\omega$ where the entanglement is largest, and we decouple it from the atom by projecting the atom onto the state $\ketbra{+}{+}$, so as to not destroy the oscillator superposition.
The Wigner function of the oscillator in this state can be written in dimensionless quadrature operators as
\begin{equation}
W_{\rho_{\mathrm{th}}}(Q,P)= W_{+}(Q,P)+W_{-}(Q,P)+W_{\text{int}}(Q,P),\label{eq:wig}
\end{equation}
where
\begin{subequations}
\begin{align}
&W_{\pm}(Q,P) = \frac{1}{N} \exp\left(\frac{1}{2\tilde{n}+1}\left(-2P^{2}-\left(\sqrt{2}Q\pm\sqrt{8}\lambda\right)^{2}\right)\right), \\
&W_{\text{int}}(Q,P) = \frac{2}{N}\exp\left(\frac{1}{2\tilde{n}+1}\left(-2P^{2}-2Q^{2}\right)\right)\cos\left(8P\lambda\right),
\end{align}
\end{subequations}
where $\lambda = g/\omega$ and where
\begin{equation}
    N = \pi(2\tilde{n}+1)\left(1+e^{-8\lambda^{2}(2\tilde{n}+1)}\right).
\end{equation}
We see that when $\lambda=0$, the Wigner function is a Gaussian centred at the origin in phase space corresponding to the initial thermal state.

The integral \eqref{eq:wigneg} must be performed numerically, and we show the results in Fig.~\ref{fig:wigneg}. We see that the negativity increased with coupling strength $\lambda$, and when the interaction strength is zero the oscillator remains in a thermal state and hence is classical. For larger $\lambda$, the oscillator state contains more coherence at a half mechanical period and hence larger negativity. However the negativity decreases with initial oscillator temperature, and to achieve large negativity with high temperature requires a larger $\lambda$ to introduce enough coherence to compensate for the thermal noise.

This can be seen directly by rewriting \eqref{eq:wig} more compactly as
\begin{align}
&W_{\rho_{\mathrm{th}}}(Q,P)\label{eq:wigcompact}  \\
&\quad= \frac{2}{N}\exp\left(\frac{-2P^{2}-2Q^{2}}{2\tilde{n}+1}\right) \nonumber \\
&\qquad\times\left(\exp\left(\frac{-8\lambda^{2}}{2\tilde{n}+1}\right)\cosh\left(\frac{8Q\lambda}{2\tilde{n}+1}\right)+\cos(8P\lambda)\right),\nonumber
\end{align}
in which the negativity of the Wigner function is caused solely by the cosine term. This term will cause more negativity with larger $\lambda$. Decreasing $\tilde{n}$ also increases the negativity as it suppresses the exp-cosh term. The Wigner function will only be negative for $Q$ sufficiently close to zero, with the troughs occurring at $P=(2n+1)\pi/8\lambda$ for $n\in \mathbb{Z}$.

In other words, the evolved state \eqref{eq:evolve} of the oscillator becomes more classical as the initial temperature increases.
This is to be expected from our first semi-classical model (see section~\ref{sec:ex1}) which identically reproduces the same visibility decline and revival as the fully quantum mechanical model in the case of high temperatures, and differs only at lower temperatures. Unless the oscillator is very close to its ground state, the non-classicality is vanishingly small, especially for very low coupling strengths. Since $\lambda \ll 1$ in Ref. \cite{Carney2021}, one would need to operate at and independently verify the ground state of the oscillator to infer entanglement generation.

Let us look more closely at the negativity in the small coupling regime $\lambda \ll 1$. Let us assume that the system is in the ground state, $\tilde{n}=0$, since as we saw earlier, the negativity decreased with increasing temperature.
From \eqref{eq:wigcompact} one can use the triangle inequality to obtain
\begin{align}
|W_{\rho_{\mathrm{th}}}(Q,P)| &\leq \frac{2}{N}\exp\left(-2P^{2}-2Q^{2}\right) \\
&\quad\times\left(\exp\left(-8\lambda^{2}\right)\cosh\left(8Q\lambda\right)+1\right).\nonumber
\end{align}
Therefore, after integrating we have
\begin{equation}
    \delta(W_{\rho_{\text{th}}}) \leq \tanh\left(4\lambda^{2}\right).
\end{equation}
For $\lambda \ll 1$, we can approximate this as
\begin{equation}
    \delta(W_{\rho_{\text{th}}}) \lesssim 4\lambda^{2}.
\end{equation}
Thus we see that for small $\lambda$ the negativity of the oscillator state produced in the experiment \eqref{eq:evolve} becomes vanishingly small, even when the system begins in the ground state. For finite temperature, it effectively vanishes.

\section{Discussion}

The authors of \cite{Carney2021} support their claim that the collapse and revival of the interferometric visibility is a true signature of entanglement by proving a theorem that shows that if no quantum entanglement is generated, the visibility cannot revive. The theorem rests on some  assumptions, and since we claim to be providing semi-classical models which do not generate any entanglement but nevertheless display the same collapse and revival signature, we ought to discuss how our models contradict the theorem proved in \cite{Carney2021}.
Here we quote the theorem in full.

\begin{theorem}\label{thm:cmt}
Let $L$ be a channel on $H_{A}\otimes H_{B}$ where $H_{A}$ is a two-state system and $H_{B}$ is arbitrary. Assume that:
\begin{enumerate}
    \item The channel $L$ generates time evolution, in a manner consistent with the time-translation invariance, thus obeying a semigroup composition law $L_{t\rightarrow t^{\prime\prime}} = L_{t\rightarrow t^{\prime}}L_{t^{\prime}\rightarrow t^{\prime\prime}}$ for all $t\leq t^{\prime}\leq t^{\prime\prime}$.
    \item The two-level subsystem $H_{A}$ has its populations preserved under the time evolution, $\sigma_{z}(t) = \sigma_{z}(0)$.
    \item $L$ is a separable channel: all of its Kraus operators are simple products. In particular, this means that any initial separable (non-entangled) state evolves to a separable state: $\rho(t) = L_{t}[\rho(0)]$ is separable for all separable initial states $\rho(0)$.
\end{enumerate}
Then the visibility $V(t)$ is a monotonic function of time.
\end{theorem}

In our semi-classical models, the quantum channel is the random unitary channel given by Eqs. \eqref{eq:scState} and \eqref{eq:scU}. Since it commutes with $\sigma_{z}$ it clearly satisfies assumption 2, and by appending an arbitrary Hilbert space $H_{B}$, then we see that $L \otimes \mathbb{I}$ satisfies assumption 3 as well. Thus the conflict with this theorem must lie in the first assumption. Indeed, the proof of Theorem 1 in Ref.~\cite{Carney2021} relies on the form of Lindblad master equation, where the Lindblad operators are time independent. The sufficient and necessary condition for the existence of such a Lindblad master equation is the divisibility of the master equation (the first condition of theorem~\ref{thm:cmt}). The additional assumption of time-translation invariance implies that the Lindblad operators are time-independent and this is used in the proof. This is equivalent to requiring the quantum channel satisfy a one-parameter semigroup composition law $L_{t_{1}}L_{t_{2}} = L_{t_{1}+t_{2}}$.

Our semi-classical models are not divisible, thus they do not correspond to a Lindblad master equation for the atom.
However our first semiclassical model is in fact time translation invariant. 
Although there is explicit time dependence in the Hamiltonian in our semi-classical models, which indicates that the Hamiltonian is not time translation invariant, that does not mean that the corresponding random unitary channel after averaging over the classical randomness is not time translation invariant. 
It is straightforward to check that our first semi-classical model is time translation invariant by showing that $L_{0\rightarrow \tau}=L_{t\rightarrow t+\tau}$ for all $t$.

The existence of our semi-classical models demonstrates that the conditions of this theorem are very restrictive, as many simple semi-classical models reproduce the decline and revival of interferometric visibility.

\section{Conclusions}
Experiments to probe the quantum nature of gravity have become promising research directions in recent years. While some proposals aim to test specific models \cite{marshall2003towards, pikovski2012probing, bekenstein2012tabletop, bassi2017gravitational}, others focus on indirect signatures of the quantization of gravity through gravitational generation of entanglement \cite{bose2017spin, marletto2017gravitationally, pedernales2021enhancing, Carney2021}. The proposed signature in Ref. \cite{Carney2021} is the loss and revival of visibility in an atomic interferometer, under specific assumptions on the dynamics. Here we show that this signature appears also in simple semi-classical models, thus such a signature by itself cannot indicate entanglement between the systems. According to the Ehrenfest theorem, the average behavior of a harmonic oscillator can be classically described. It is thus important to explore the classical picture to test whether specific signatures can reflect quantum behavior.
The semi-classical models we have discussed are reasonably general and applicable to other systems and states. In fact, a random unitary channel represents the time evolution of a quantum system under the influence of classical systems containing classical uncertainties~\cite{audenaert2008random}. As the collapse and revival of the interference visibility can be reproduced by the atom subject to random unitary channels, they cannot be considered good signatures of entanglement generation.

Our results do not contradict the claims of Ref. \cite{Carney2021} since the models we present do not satisfy the conditions on the dynamics under which entanglement can be inferred. But our findings highlight that such conditions are violated in many semi-classical scenarios. This indicates that the conditions imposed on the dynamics in Ref. \cite{Carney2021} are very restrictive: they exclude simple and reasonable classical dynamics, and thus leave little room to test against the inference of entanglement generation unless there is supplementary evidence that the conditions are satisfied. For the very low coupling strengths envisioned in the experiment, the non-classicality vanishes unless the oscillator is nearly exactly in its ground state.
It therefore remains a significant challenge to verify the quantum nature of the interaction in such an experimental scenario. 

\emph{Note added} -- We were made aware of Ref.~\cite{hosten2021testing} after we complete the manuscript. Ref.~\cite{hosten2021testing} is closely related to Sec.~\ref{sec:ex1} to show that collapses and revivals can also be explained using a semi-classical mean-field model.


\acknowledgements

MSK and YM thank EPSRC for financial supports (EP/R044082/1, EP/P510257/1). IP and TG acknowledge support by the Swedish Research Council under
grant no. 2019-05615. IP also acknowledges support by the European Research Council under grant no. 742104 and The Branco Weiss Fellowship -- Society in Science.

\providecommand{\noopsort}[1]{}\providecommand{\singleletter}[1]{#1}%

\end{document}